\documentclass[prb,aps,amsmath,amssymb,floatfix,prb,twocolumn]{revtex4}

\usepackage{epsfig}
\begin{document}

\title{Nonlinear ac  conductivity of one-dimensional Mott insulators}

\author{B.~Rosenow}\thanks{On leave from the Institut f\"ur Theoretische Physik, Universit\"at zu
K\"oln, D-50923, Germany}
\affiliation{Physics Department, Harvard University, Cambridge, MA
02138, USA}

\date{February 20, 2008}

\begin{abstract}

  We discuss a semiclassical calculation of low energy charge
  transport in one-dimensional (1d) insulators with a focus on Mott
  insulators, whose charge degrees of freedom are gapped due to the
  combination of short range interactions and a periodic lattice
  potential.  Combining RG and instanton methods, we calculate the
  nonlinear ac conductivity and interpret the result in terms of
  multi-photon absorption.  We compare the result of the semiclassical
  calculation for interacting systems to a perturbative, fully quantum
  mechanical calculation of multi-photon absorption in a 1d band
  insulator and find good agreement when the number of simultaneously
  absorbed photons is large.

\end{abstract}


\maketitle

%%%%%%%%%%%%%%%%%%%%%%%%%%%%%%%%%%%%%%%%%%%%%%%%%%%%%%%%%%%%%%%%%%%%
\section{Introduction}
%%%%%%%%%%%%%%%%%%%%%%%%%%%%%%%%%%%%%%%%%%%%%%%%%%%%%%%%%%%%%%%%%%%%

In 1d electron systems,  interactions have a very pronounced effect and give 
rise to a variety of 
unusual phenomena  \cite{Giamarchi,Gruner}. Interacting 1d systems with a lattice potential display insulating behavior  at commensurate filling. Examples include charge density waves (CDWs), and fermions
 on a lattice with local Hubbard interactions. Due to interaction effects, the spin degrees of freedom are gapped and decoupled from the charge degrees of freedom, such that they do not influence the dynamics of charge transport. 
In these systems, the charge degrees of freedom are described  by the quantum sine-Gordon model.

The low frequency response of a half filled system with periodic potential
is characterized by an optical gap. Hence at zero temperature  the linear ac
conductivity vanishes for frequencies smaller than the gap, apart from
a series of discrete resonances which appear for sufficiently strong
repulsive interaction \cite{KlMa79,CoEsTs01}. For frequencies larger
than the gap energy the linear conductivity shows a power law
dependence on frequency \cite{Giamarchi91}.  Similarly, 
the linear dc conductivity shows a power law
dependence on temperature $T$ at  temperatures
larger than the spectral gap.

 At zero
temperature and frequency, charge transport is only possible by
tunneling of charge carriers, which can be described by instanton
formation. The nonlinear dc-conductivity is characterized by\cite{Maki77,Maki78} $I \sim
\exp(- E_0 / E)$, and for fixed value of 
the electric field strength is a monotonically increasing function of the ac frequency of the external field. In this work, we revisit the instanton calculation of the   nonlinear ac
  conductivity by Maki and  discuss in detail the physics of multi-photon absorption in the dynamic limit. In addition, we point out that in 
 the static limit  the efficiency of soliton-antisoliton pair annihilation and the 
 dynamics of scattering between these quasiparticles decisively determines
 the value of the conductivity.

 To be specific, we consider the quantum sine-Gordon model. 
We first scale the
system to its correlation length, where the influence of the potential
is strong and a semiclassical instanton calculation becomes
possible.  We reproduce  an earlier result \cite{Maki78} for the nonlinear ac current within the Matsubara formalism, and carefully analyze  
the dependence of the instanton creation rate on the ratio of electric field energy
to photon energy.   In the dynamic limit, the result can be interpreted in 
terms of multi-photon absorption. We compare the result of our semiclassical 
analysis to a fully quantum mechanical perturbative expression for 
a 1d band insulator. When expressing the semiclassical result in terms 
of the energy gap, the solitonic correlation length and effective velocity, 
it can be compared to the quantum mechanical expression. In the limit where
the photon energy is much smaller than the optical gap, there is good agreement between the two expressions.

%%%%%%%%%%%%%%%%%%%%%%%%%%%%%%%%%%
\setcounter{equation}{0}

\section{Perturbative calculation for  noninteracting electrons}

%%%%%%%%%%%%%%%%%%%%%%%%%%%%%%%%%%%

As a reference point for the instanton calculation for  interacting systems, we start by describing  a fully quantum mechanical calculation  of the conductivity of  noninteracting 1d electrons with a periodic potential at half filling. The periodic potential splits the band into a valence and a conduction band, such that the system becomes
a band insulator. When exposing the system to a monochromatic electric field
$E(t) = E_0 \cos(\omega t)$, the only contribution to the ac conductivity comes 
from the excitation of electron-hole pairs. 
Clearly, the linear ac conductivity vanishes as long as 
the photon energy $\hbar \omega$ is smaller than the band gap $E_{\rm gap}$. 
In the following, we sketch a derivation of the {\em nonlinear} ac conductivity.

We follow the derivation described by Wherrett \cite{Whe84}. We consider a situation in which $N$ photons are needed to excite an electron from the valence to the conduction band, that is $N \hbar \omega > E_{\rm gap}$.  
The coupling to the electromagnetic field has its origin in the $A\cdot p$
term in the Hamiltonian, its matrix elements are given by
%
%************************  matrix elements of electromagnetic coupling ****
\begin{equation}
(\hat{\cal V})_{in} \ = \ {e_0 E_0 \over i \omega m_{\rm eff}} \ (\hat{p})_{i n}  \ 
\approx \ {e_0 E_0 v_{\rm eff} \over i \omega} 
\end{equation}
%********************************************************
%
In the last step it was assumed that the initial and final momentum are close to the Fermi momentum, and that the ratio $p_F /m_{\rm eff} = v_{\rm eff}$ defines an effective velocity. 
We calculate the transition rate from Fermi's golden rule as
%
%********* golden rule transition rate  *************************
\begin{equation}
P_N(E_0, \omega) \ = \ {2 \pi \over \hbar} |M^{(N)}_{vc} |^2 \rho(N \hbar \omega - E_g) \ \ .
\end{equation}
%******************************************************
%
The  density of states is given by $\rho(\epsilon) = (\sqrt{\hbar k_F v_{\rm eff}/\epsilon})/(2 \sqrt{2} \pi 
\hbar v_{\rm eff})$.

The transition amplitude $M^{(N)}_{vc}$ between valence and conduction band
involves $N-1$ virtual intermediate states. For odd $N$, these states correspond to the consecutive excitation and deexcitation between valence and 
conduction band. Assuming that the transition matrix elements have no important energy dependence, it can be calculated as
%
%********************  effective transition matrix element **************
\begin{eqnarray}
|M^{(N)}_{vc}| & = & \left({e_0 E_0 v_{\rm eff} \over \omega} \right)^N\ \  \prod_{j=1}^{N-1}
{1 \over E_g - 2 j \hbar \omega} \nonumber \\
& \approx & {\hbar \omega  \over  \pi} \  \left({e_0 E_0 v_F  \over  \omega E_g  }{e \over 2} \right)^N   \  .
\end{eqnarray}
%*****************************************************************
%
In the last step, it was assumed that $E_g \approx N \hbar \omega$ and Stirling's formula for the factorial function was used. Putting everything together, 
the transition rate per unit length is found to be
%
%*********************************  transition rate *************
\begin{eqnarray}
P_N(E_0, \omega) & = &  {\hbar \omega^2 \over \sqrt{2} \pi^2 v_{\rm eff}} \ 
 \left({e_0 E_0 v_{\rm eff}  \over  \omega E_g  }{e \over 2} \right)^{2 N}  \sqrt{\hbar k_F v_{\rm eff} \over 
 N \hbar \omega - E_g} \nonumber \\
 &  &\cdot  \Theta(N \hbar \omega - E_g) \label{perturbative.eq}
\end{eqnarray}
%*************************************************
From this absorption rate, the $N$-photon contribution to the nonlinear conductivity can be obtained via
 %
%******************* conductivity and energy absorption ***********
\begin{equation}
\sigma_N(\omega,E_0) \ = \ {2 N \hbar \omega\ \over E_0^2} 
P_N(E_0, \omega)  \ \ . \label{absorption.eq}
\end{equation}
%*****************************************************************
%

%%%%%%%%%%%%%%%%%%%%%%%%%%%%%%%%%%
\setcounter{equation}{0}

\section{Model}

%%%%%%%%%%%%%%%%%%%%%%%%%%%%%%%%%%%

The charge degrees of freedom of interacting 1d electrons subject to a periodic potential are described by the quantum sine-Gordon model
%
%***************************** action **********************************
\begin{eqnarray}
{S\over \hbar} & = & { 1 \over 2 \pi K} \int dx \int_0^{v  \hbar \beta} d y
\left[ \left({\partial \varphi \over \partial x} \right)^2 +  \left({\partial \varphi
\over \partial y}\right)^2
\right.
\label{action.eq}\\
& & \left. - 2 u \cos( p \varphi ) + {2 K e_0 \over \pi  v} \varphi  E(y)
\right] \ + {S_{\rm diss} \over \hbar} ,
\nonumber
\end{eqnarray}
%******************************************************************
%
where we have rescaled time according to $v \tau \to y$, and 
$\beta = 1/k_B T$.  The
dissipative part of the action describes a weak coupling of the electron
system to a dissipative bath, for example phonons. It is needed for
energy relaxation in soliton-antisoliton annihilation processes 
\cite{Buttiker86}. We assume it to be so small that it does not 
 influence the RG equations for the other model parameters significantly.
The
smooth part of the density is given by ${1 \over \pi} \partial_x
\varphi$, and $p=1,2$ for CDWs and LLs, respectively. 

For $K > K_c(u)$ the potential is RG irrelevant and decays under the
RG flow, while for $K < K_c(u)$ the potential is relevant and grows.  For the 
periodic potential in the action Eq.~(\ref{action.eq}), one finds 
 $K_c(0) = 8/p^2$.     We assume $K < K_c(u)$
and scale the system to a length $\xi= a e^{l^\ast}$, on which the
potential is strong.  After the scaling process, the parameters $K$,
$v$, and $u$ in Eq.~(\ref{action.eq}) are replaced by the effective,
i.e. renormalized but not rescaled, parameters $K_{\rm eff}$, $v_{\rm
  eff}$, and $u_{\rm eff}$.

 The  compressibility $\kappa = {\partial \rho \over \partial \mu}$ is used
 as a generalized density of states for interacting systems, it is given by
 $\kappa_{\rm eff}=K_{\rm eff}/v_{\rm eff} \pi$. Our calculations are valid for photon energies  $\hbar \omega$ and electric field energies    
 $e E_0 \xi K_{\rm eff}$ below the soliton energy
$ E_s = { 8 \over \kappa_{\rm eff} \xi \pi^2  p^2}$.  In this RG calculation, we do not attempt to treat a possible nonlinear 
 dependence of coupling parameters on the external electric field. 
 The full inclusion of the external field in an equilibrium theory is not possible
 as it renders the ground state of the system  unstable. 
 The quantum sine-Gordon model has an infinite number of 
 ground states connected by a shift of the phase field by $2 \pi $.
 Here, we concentrate on renormalizing each of these 
  ground states 
 separately and  take into account the coupling between different ground 
 states due to the external electric field in the framework of an instanton 
 approach.

%%%%%%%%%%%%%%%%%%%%%%%%%%%%%%%%%%%%
\setcounter{equation}{0}

\section{Instantons}
%%%%%%%%%%%%%%%%%%%%%%%%%%%%%%%%%%%%

We consider a time dependent external field
$E(t)$, which upon analytical continuation $it v_{\rm eff} \to y$ turns into 
a field $E(y)$. In imaginary time, the electric field  has to obey the same
 periodic boundary 
condition $E(y + \beta v_{\rm eff}) = E(y)$ as other bosonic fields, e.g.
the displacement field $\varphi(y)$.
This boundary condition is respected by a discrete Fourier
representation \cite{NeOr88}
%
%***********************  Matsubara sum ****************************
\begin{equation}
E(y) = T \sum_{\omega_n} E(\omega_n)
e^{- i \omega_n y/v_{\rm eff}}, \ \ \    \omega_n = 
{n 2 \pi k_B T \over \hbar}
\label{matsubara.eq}
\end{equation}
%********************************************************************
with Matsubara frequencies $\omega_n$. A monochromatic external 
field is hence described by $E(y) = E_0 \cos (\omega_n y/v_{\rm eff})$.
Analytic continuation from  Matsubara frequencies to  real frequencies
is defined by
$i \omega_n \to \omega + i \eta$.
The original calculation \cite{Maki78} did not make use of Matsubara frequencies and used a dependence on imaginary time 
%
%*****************  time dependence Maki  *********************
\begin{equation}
 E(y) \ = \ E_0 \cosh \left(\omega y/v_{\rm eff}\right)\ \ , \label{cosh.eq}
 \end{equation}
 %**************************************************
 %
  which does not respect the periodicity requirement in imaginary 
time and grows exponentially in $y$. We define the ratio
% 
%***********************  gamma definition ***************************
\begin{equation} 
\gamma= {e_0 E_0 \xi  K_{\rm eff} p \over \hbar \omega }\  {  \pi 
  \over 4} \ \ ,
\end{equation}
%****************************************************************
%
which is proportional to the ratio of the field energy acquired on a length 
scale $K_{\rm eff} \xi$ and the photon energy $\hbar \omega$.
The static dc limit corresponds to $\gamma \to \infty$, whereas the dynamic ac limit corresponds to $\gamma \to 0$. It turns out that for $\gamma > 1$ it does not 
matter whether one analytically continues to real frequencies before solving the 
equation of motion for the instanton or in the very end of the calculation. 
However, in the range  $\gamma  < 1$, the two prescriptions lead to different results. Using an oscillatory time dependence as defined by 
a monochromatic $E(\omega_n)$ in Eq.~(\ref{matsubara.eq}), the instanton 
solution becomes periodic in time and has no well defined beginning or end. 
On the other hand,  using the time dependence Eq.~(\ref{cosh.eq}) in the equation of 
motion, the instanton is well defined and  the final result agrees quantitatively with the quantum mechanical
calculation presented in the previous section. Although it is not very intuitive why the exponential time dependence Eq.~(\ref{cosh.eq}) should describe an
electric field periodic in time, it leads to the correct result and we adopt this 
prescription in the following calculation.

The wall width $1/\sqrt{ u_{\rm eff}} \approx \xi$ of an instanton
solution to the action Eq.~(\ref{action.eq}) is for weak external fields
much smaller than the extension $ E_s/e_0 E_0$ of the instanton.
The domain wall energy per unit length is given by $E_s/v_{\rm eff}$, 
and in contrast to the disordered case \cite{RoNa06},  the spontaneous 
formation of instantons due to quantum fluctuations needs not be considered 
as the domain wall energy is spatially constant  in the present model.
Hence, the instanton action can be expressed in terms of the domain
wall position $X(y)$ as
%
%****************** domain wall action *******************************
\begin{eqnarray}
{S_{\rm wall}\over \hbar} & = &   {E_s \over v_{\rm eff} }     \oint dy
\sqrt{1 + (\partial_y X)^2} \nonumber \\ 
& - & {2 e_0 \over 
  v_{\rm eff} p} \int \!  d y E(y) [ X_+(y)\!  -\!  X_-(y)]
\
\label{wallaction.eq}
\end{eqnarray}
%**********************************************************************
%
The
equation of motion can be integrated to give
%
%**************************** general solution ************************
\begin{equation}
X_{\pm}(y) = \mp \gamma \int_{y_0}^y d z { \sinh \left(\omega z/v_{\rm eff} \right)
\over \sqrt{ 1 - \gamma^2 \sinh^2 \left(\omega z/v_{\rm eff}\right)}} \ \ .
\label{wallequation.eq}
\end{equation}
%*********************************************************************
%
% 
%
%**************************** figure sine gordon instantons *************
\begin{figure}[t]
%\vspace*{-1.5cm}
\centerline{ \epsfig{file=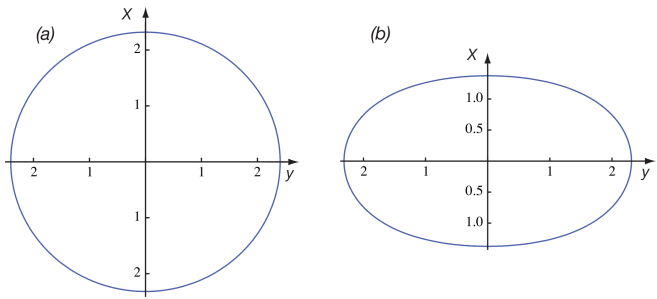,width=\linewidth}}
\caption{Instantons  for different
  ratios $\gamma= {e_0 E_0 \xi  K_{\rm eff}  \over \hbar \omega } {p \pi 
  \over 4} $ of field
  strength to frequency. (a) for $\gamma 21$, the instanton is almost circular, 
  and  (b) for $\gamma = 0.2$, the
  instanton is elongated in time direction.}
\label{instantons.fig}
\end{figure}
%***********************************************************************
%
The
domain of integration is bounded by the singularities $\pm y_0$ of the
integrand and the solution corresponds indeed to an instanton with
finite Euclidean action (Fig.~\ref{instantons.fig}).  The spatial extension of the instanton is 
%
%*********************  space-time instanton extension  ***************
\begin{eqnarray}
L_x(\omega, \gamma)  & = &  {2 v_{\rm eff} \over \omega} \ \arctan{1 \over \gamma}
\nonumber \\
L_t(\omega, \gamma) & = & {2 \over \omega} \ {\rm arcsinh} {1 \over \gamma} \ \ .
\label{spacetimeextension.eq}
\end{eqnarray}
%******************************************************************
%
The limiting forms for the static limit $\gamma \to \infty$ and the dynamic limit
$\gamma \to 0$ are
%
%***********  space time instanton extension limiting forms ***********
\begin{eqnarray}
L_x(\omega, \gamma)  & = &  v_{\rm eff} \ \left\{ \begin{array}{cc}  {8 \hbar v_{\rm eff} \over
e_0 E_0 \xi K_{\rm eff} p \pi} &  \gamma \to \infty \\[.5cm]
{\pi v_{\rm eff} \over \omega} & \gamma \to 0
\end{array} \right.\\[.5cm]
L_t(\omega, \gamma)  & = &   \left\{ \begin{array}{cc}  
{8 \hbar  \over
e_0 E_0 \xi K_{\rm eff} p \pi} &  \gamma \to \infty \\[.5cm]
{2 \ln{2\over \gamma} \over \omega} & \gamma \to 0
\end{array} \right.
\end{eqnarray}
%*************************************************************
%
In the static limit, the instanton is circular with an extension inversely proportional to the electric field. In the dynamic limit, the scale for the instanton extension is set by the period of the ac electric field,  and  it is elongated in time direction by a factor $\ln(2/\gamma)$. 

\begin{figure}[t]
\centerline{ \epsfig{file=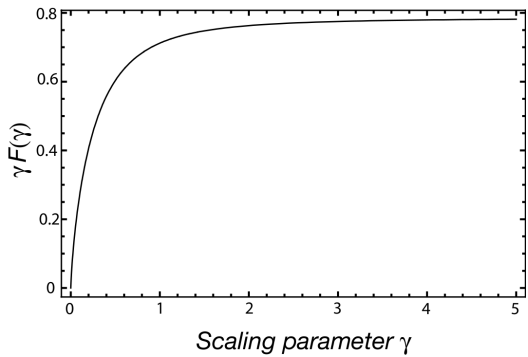,width=7cm}}
\caption{ Dependence of the exponent of $P_E$ on   $\gamma= {e_0 E_0 \xi  K_{\rm eff}  \over \hbar \omega }\  { \pi 
  \over 4}$. The exponent grows monotonically with $\gamma$ and saturates 
  for $\gamma \to \infty$.}
\label{elliptic.fig}
\end{figure}

The creation rate of kink antikink pairs is proportional
to the negative exponential of the instanton action divided by the space-time
volume of an instanton. One finds
%
%*********************** current equation **************************
\begin{eqnarray}
P(E_0, \omega) & =  &   { \exp{ \left[- { 2 E_s \over  e_0 E_0 \xi K_{\rm eff} p}\; {4 \over \pi} \;  \gamma \;
{\rm  F}(\gamma) \right]}\over L_x(\omega, \gamma) L_t(\omega, 
\gamma) }      
\label{creationrate.eq} \\[0.5cm]
 {\rm  F}(\gamma)&  =  & \int_0^{{\rm arsinh} {1 \over \gamma} }
\! \! \! \! \! \! \! \! \! du
\sqrt{1 - \gamma^2 \sinh^2 u}\ .
\nonumber
 \end{eqnarray}
 %***************************************************************
 % 
 This result would be changed by a preexponential factor when including the 
 contribution of 
fluctuations of the instanton shape, of zero modes, and of the Jacobi
determinant originating in the transition from an integral over the
field $\varphi$ to an integral over the the domain wall position 
\cite{Langer67,Kleinert}.  The dependence of the exponent on $\gamma$
is displayed in Fig.~\ref{elliptic.fig}. For the static limit with frequencies  $\omega \ll e_0 E_0 \xi$, $\gamma$ is large 
and we find 
%
%******************* rate for small frequencies ********************
\begin{equation}
 \gamma F(\gamma)\to {\pi \over 4} \left[ 1 - {1 \over 8 \gamma^2} +
 O({1 \over \gamma^4}) \right] \ \ .
 \label{correction.eq}
 \end{equation}
 %*********************************************************************
 %
According to the expansion Eq.~(\ref{correction.eq}), the first correction of 
order $\omega^2$ reduces the static value of the exponent. 
Hence, for a fixed field amplitude $E_0$, the creation rate increases 
monotonically with frequency. 

In the dynamic  limit of large frequencies, one finds
%
%******** rate for large frequencies  ************************
\begin{equation}
\gamma F(\gamma) \ \to \ \gamma \ln \left( {4 \over e \gamma} \right) \ \ .
\end{equation}
%******************************************************
%
This result gives rise to an instanton creation rate
%
%*************  instanton creation rate  ************
\begin{equation}
P(E_0,\omega) =  {\omega^2 \over 2 \pi v_{\rm eff} \ln{2 \over \gamma}}    \left( {e_0 E_0 K_{\rm eff} \xi p\over \hbar \omega} 
{e \pi \over 16} \right)^{4 E_s \over \hbar \omega}  \ \ .
\end{equation}
%***************************************************
%
This instanton creation rate can be interpreted  in terms of multi-photon absorption:
the optical gap in the system is $ 2 E_s$, and hence the smallest integer larger than  the ratio
$2 E_s \over \hbar \omega$ determines the number of photons necessary to create one soliton-antisoliton pair. The ratio of electric field strength to photon energy 
in the brackets agrees exactly with the noninteracting result 
Eq.~(\ref{perturbative.eq}) if one identifies $E_g = 2 E_s$, 
uses the definition of the soliton energy, and considers the fermionic case
 $p=2$.

The present  semiclassical approximation 
can be expected to work well for $\hbar \omega \ll E_s$ when the photon 
number is large and the electrical field can be treated unquantized. However, 
it does not reproduce the threshold behavior expected whenever the photon energy is tuned through an integer fraction of the optical gap $2 E_s$.
The specific physics of a Mott insulator as compared to a band insulator is 
 contained in the energy dependent form factor, which for the band 
insulator is just the density of states. The energy dependence of the form factor 
is probably contained in fluctuation corrections around the saddle point, which 
would be interesting to calculate.  The main conclusion from the agreement between the semiclassical and the quantum mechanical result is that the dominant electric 
field dependence for the Mott insulator is similar to that of the band insulator 
when expressed in suitable effective quantities, which parameterize the 
interaction strength.

%         
%
%**************************** figure gamma dependence *************
%***********************************************************************
 %
 
 %%%%%%%%%%%%%%%%%%%%%%%%%%%%%%%%%%%%
\setcounter{equation}{0}

\section{Nonlinear Conductivity}
%%%%%%%%%%%%%%%%%%%%%%%%%%%%%%%%%%%%
 
The quantum sine-Gordon model is integrable, and in the absence of additional 
interaction terms or a coupling to a dissipative bath its solitonic excitations 
have an infinite life time. According to this logic, even an infinitesimally small 
density of solitons would give rise to an infinite dc conductivity. This 
conclusion does not sound realistic, and in the following we will discuss
the behavior expected from physical systems which are approximately 
described by this model.

As both kinks and antikinks carry a
topological quantum number, they cannot simply decay but only
annihilate each other \cite{Buttiker86}. Pairwise annihilation gives
rise to a decay rate $P_{\rm decay} \sim n_{\rm eq}^2$ and hence $n
\sim \sqrt{P(E_0,\omega)}$ in equilibrium. As kinks and antikinks are solitary
waves, this annihilation is only possible in the presence of a
dissipative bath. 

In the dynamic limit with multi-photon absorption, the kink and anti-kink created by the electric field stay close to each other and it seems plausible that they 
will annihilate quickly such that the nonlinear ac conductivity can be 
calculated according to Eq.~(\ref{absorption.eq}). 

The static limit is more subtle. Maki \cite{Maki77,Maki78} suggested 
to calculate the conductivity according to  Eq.~(\ref{absorption.eq}) in the dc limit as 
well, which would imply that every kink and antikink that come close 
to each other annihilate with certainty. This assumption would only be 
justified if the coupling to the dissipative bath was strong.  If the coupling to the 
dissipative bath was weak, there would be a finite density of kinks and antikinks, and the conductivity would depend on their mean free path.
Under the assumption \cite{DaSa05} that the mean free path is 
of the order of the inter-particle distance, the product of equilibrium density 
and mean free path would be of order one, and the conductivity would 
not be exponentially suppressed anymore,  a result quite different from 
 the exponential suppression in the case of strong dissipation.

In summary, we have discussed the  nonlinear ac conductivity
of interacting 1d electron systems in a periodic potential. We found that 
the dominant physical process in the dynamical limit is multi-photon absorption, and that the 
result of an instanton calculation is in quantitative agreement with 
a fully quantum mechanical calculation for noninteracting electrons in the 
limit of a large number of simultaneously absorbed photons.

{\em Acknowledgments:}  This paper is dedicated to Thomas Nattermann on 
the occasion of his 60th birthday. I would like to thank him for generously sharing his deep knowledge about disordered systems with me, and I am 
indebted to him  for collaboration in an early stage of this research. I would like to  thank  E.~Heller  and S.~Sachdev for interesting  discussions, and the Heisenberg program of DFG for support.

\end{document}